# Rogue waves for a system of coupled derivative nonlinear Schrödinger equations

H. N. Chan[(1)], B. A. Malomed[(2)], K. W. Chow[(1)]*, E. Ding[(3)]

[(1)] Department of Mechanical Engineering

University of Hong Kong, Pokfulam, Hong Kong

[(2)] Department of Physical Electronics, School of Electrical Engineering,

Tel Aviv University, Tel Aviv 69978, Israel

[(3)] Department of Mathematics and Physics, Azusa Pacific University,

Azusa, CA 91702, USA

* K. W. Chow (Corresponding author)

Fax: (852) 2858 5415   Email: kwchow@hku.hk

# ABSTRACT

Rogue waves (RWs) are unexpectedly strong excitations emerging from an otherwise tranquil background. The nonlinear Schrödinger equation (NLSE), a ubiquitous model with wide applications to fluid mechanics, optics and plasmas, exhibits RWs only in the regime of modulation instability (MI) of the background. For system of multiple waveguides, the governing coupled NLSEs can produce regimes of MI and RWs, even if each component has dispersion and cubic nonlinearity of opposite signs. A similar effect will be demonstrated for a system of coupled derivative NLSEs (DNLSEs), where the special feature is the nonlinear self-steepening of narrow pulses. More precisely, these additional regimes of MI and RWs for coupled DNLSEs will depend on the mismatch in group velocities between the components, as well as the parameters for cubic nonlinearity and self-steepening. RWs considered in this work differ from those of the NLSEs in terms of the amplification ratio and criteria of existence. Applications to optics and plasma physics are discussed.

# I. INTRODUCTION

Rogue waves (RWs) are surprisingly tall localized excitations which spontaneously develop on top of an otherwise calm background [1]. Although such "killer" waves were known to the maritime community for a long time [2], the recent interest was ignited by the experimental observations of such waves in optical waveguides [3]. At present, RWs constitute subjects of intensive studies across a broad spectrum of physical disciplines.

One ubiquitous model for wave propagation in dispersive media is the nonlinear Schrödinger equation (NLSE) [4, 5]. The Peregrine soliton, a spatiotemporally localized and exact algebraic expression arising from the long wavelength limit of a pulsating mode (breather) [6], is a popular solution of the NLSE to model a RW. Theoretically this RW solution is only nonsingular when dispersion and nonlinearity of the NLSE are of the same sign. This is a regime admitting modulation instability (MI) of the plane (or continuous) wave background, and MI provides a physical mechanism to drive the RWs.

For dispersive media allowing co-propagation of multiple wave packets, a natural extension is a system of coupled NLSEs. The study of RWs in such systems chiefly focuses on the case of equal self-phase modulation (SPM) and cross-phase modulation (XPM) coefficients, known as the integrable Manakov system [7]. For the case when the dispersion and nonlinearity have the same sign in each component of the system, the RWs have been studied extensively [8 – 11].

For configurations where the dispersion and nonlinearity are of opposite signs for each component, coupling may nevertheless still give rise to MI. Although this phenomenon was discovered in such coupled NLSEs more than twenty years ago [12], recent works have demonstrated that the long wavelength (zero frequency) limit of such MI is closely related to the existence of RWs [13, 14].

In many applications, realistic models of such NLSEs should include additional nonlinear effects. Among such extended equations, important examples are the derivative NLSEs (DNLSEs), with the integrable Kaup-Newell [15] and Chen-Lee-Liu equations [16] being intensively studied representatives. Besides the technique of inverse scattering transform [15], the Hirota bilinear method is also applicable to these equations [17]. In addition to their great significance as examples of nonlinear dynamics in a general setting, these equations are also well known as relevant models for the properties of very narrow pulses in nonlinear optics [18, 19], the propagation of Alfvén waves in magnetized plasmas [20, 21], and the description of electromagnetic waves in an antiferromagnetic medium [22].

The objective of the present work is to study the dynamics of RWs in systems of coupled DNLSEs. In this context, the term 'rogue wave' refers loosely to an entity localized in both space (*x*) and time (*t*). The robustness and 'structural stability' are illustrated through numerical simulations of the wave profiles subject to random perturbations. The natural question to ask is whether coupling will produce MI regimes, while each component in isolation does not feature such an instability. The answer is affirmative. Similar to the coupled NLSEs, the existence of RWs will be shown to coincide with the presence of the MI in the long wavelength (zero frequency) limit.

We shall employ the Hirota bilinear transform, as this technique has been shown to work effectively for calculating multi-soliton solutions [23, 24], as well as RWs [25] of integrable nonlinear evolution equations. This method may be utilized as an alternative to the widely used Darboux transformation [5]. Indeed the bilinear method has already been applied to construct conventional solitons for coupled DNLSEs. Colliding solitons may then exhibit complete or partial energy switching among the components [26].

The rest of the paper is structured as follows. The coupled DNLSE system is introduced in Section II, and the RW solutions are derived as a low frequency limit of the breathers. The correlation of the existence of RWs with MI is elucidated in Section III. Wave profiles and dynamics are discussed in Section IV, where special attention is paid to the possibility of reaching higher than normal RW amplification ratio. A gauge transformation connecting the present system and other models studied in the literature is presented in Section V, together with a discussion on applications to optics and plasma physics. Conclusions are drawn in Section VI.

## II. THE EVOLUTION MODEL AND ROGUE WAVES

A system of coupled DNLSEs for wave amplitudes $A$ and $B$ is taken as [16, 26, 27]

$$iA_t + i\delta A_x + A_{xx} - \sigma(|A|^2 + |B|^2)A + i\gamma(A^*A_x + B^*B_x)A = 0,$$

$$iB_t - i\delta B_x + B_{xx} - \sigma(|A|^2 + |B|^2)B + i\gamma(A^*A_x + B^*B_x)B = 0, \qquad (1)$$

where the asterisk stands for complex conjugation, $\delta$ represents the group-velocity mismatch between the components, $\sigma$ and $\gamma$ measure the strengths of cubic nonlinearity and self-steepening effect respectively. Eq. (1) can be mapped to a more familiar form through a gauge transformation as shown later in Section V. The main goal here is to study the effect of the interplay among these ingredients on the dynamics of RWs.

The RW solution of Eq. (1) will be obtained by the Hirota bilinear method [23, 24]. A sequence of transformations is first implemented:

$$A = \rho \exp(i\beta x - i\omega_1 t)\frac{g}{f}, \quad B = \rho \exp(i\beta x - i\omega_2 t)\frac{h}{f}, \qquad (2)$$

where $\beta = -\frac{\gamma \rho^2}{2}$, $\omega_1 = \delta\beta - 3\beta^2 + 2\sigma\rho^2$, $\omega_2 = -\delta\beta - 3\beta^2 + 2\sigma\rho^2$, $\rho > 0$, and $g$, $h$, $f$ are complex functions. The parameters $\beta$, $\rho$, $\omega_n$ ($n = 1, 2$) stand for the wavenumber, amplitude and frequencies of the continuous wave respectively. For simplicity, we have assumed equal background amplitude for the two components $A$ and $B$. Situations with distinct amplitudes for the two waveguides will be left for future studies. On defining the Hirota bilinear derivative as [23, 24]

$$D_x^m D_t^n g \cdot f = \left(\frac{\partial}{\partial x} - \frac{\partial}{\partial x'}\right)^m \left(\frac{\partial}{\partial t} - \frac{\partial}{\partial t'}\right)^n g(x,t) f(x',t') \bigg|_{x=x', t=t'},$$

the bilinear form is given by

$$[iD_t + i(\delta + 2\beta)D_x + D_x^2] g \cdot f = 0,$$

$$[iD_t + i(-\delta + 2\beta)D_x + D_x^2] h \cdot f = 0,$$

$$(D_x^2 - 4\beta i D_x) f \cdot f^* = \frac{i\gamma\rho^2}{2}(D_x g \cdot g^* + D_x h \cdot h^*) + \sigma\rho^2 (2|f|^2 - |g|^2 - |h|^2),$$

$$D_x f \cdot f^* = \frac{i\gamma\rho^2}{2}(-2|f|^2 + |g|^2 + |h|^2). \tag{3}$$

The first two equations in Eq. (3) are complex ones, while the third and fourth equations reduce to purely real and purely imaginary forms respectively. Hence Eq. (3) is a system of six equations for six unknowns (real and imaginary parts of $f$, $g$ and $h$).

RWs are derived by taking the long wavelength limit of breather solutions of Eq. (3). The details of the derivations are similar to those outlined in earlier works [25, 28, 29] and are thus omitted here. It is convenient to scale the coefficients of the cubic nonlinearity and self-steepening by means of the amplitude ρ from Eq. (2) as follows:

$$S = \sigma\rho^2, \Gamma = \gamma\rho^2. \tag{4}$$

The RW solution for Eq. (1) is then given by Eq. (2) in combination with the following expressions for functions $f$, $g$, and $h$:

$$f = (x-at)^2 + b^2 t^2 + \frac{1}{b^2} + (x-\Omega_0 t)\frac{4\Gamma i\left[(\Omega_0^* + \Gamma)^2 + \delta^2\right]}{\left[(\Omega_0^* + \Gamma)^2 - \delta^2\right]\left[-(\Omega_0^*)^2 + \delta^2 + \Gamma^2\right]},$$

$$g = f + (x-\Omega_0 t)\frac{4i(\delta - a - \Gamma)}{(a-\delta+\Gamma)^2 + b^2} + \frac{4b}{(\Omega_0 - \delta + \Gamma)}t$$
$$+ \frac{4\left[(\Omega_0^* + \delta + \Gamma)^2 - 2\delta\Gamma\right]}{(\Omega_0 - \delta + \Gamma)(\Omega_0^* + \delta + \Gamma)\left[-(\Omega_0^*)^2 + \delta^2 + \Gamma^2\right]},$$

$$h = f + (x-\Omega_0 t)\frac{4i(-\delta - a - \Gamma)}{(a+\delta+\Gamma)^2 + b^2} + \frac{4b}{(\Omega_0 + \delta + \Gamma)}t$$
$$+ \frac{4\left[(\Omega_0^* - \delta + \Gamma)^2 + 2\delta\Gamma\right]}{(\Omega_0 + \delta + \Gamma)(\Omega_0^* - \delta + \Gamma)\left[-(\Omega_0^*)^2 + \delta^2 + \Gamma^2\right]}. \tag{5}$$

The parameters $a$, $b$ denote the real and imaginary parts of the frequency $\Omega_0$,

$$a = \text{Re}(\Omega_0), b = \text{Im}(\Omega_0), \tag{6}$$

where this complex valued frequency is the leading order approximation in the zero frequency expansion of the dispersion relation,

$$\Omega_0^4 - 2(2S + \delta^2)\Omega_0^2 + 4(\Gamma^3 - 2S\Gamma)\Omega_0 + (\Gamma^2 + \delta^2)(3\Gamma^2 - 4S + \delta^2) = 0. \tag{7}$$

Given the algebraic complexity of Eq. (7), it is not possible to perform an explicit analytical investigation. However, all numerically tested examples of the solution exhibit nonsingular algebraically localized RW patterns.

A remark on the permissible forms of complex frequencies is in order. As the cubic term is absent in Eq. (7), the sum of the roots must be zero. Consequently, if real roots are absent, complex ones must appear in complex conjugate pairs of the form of $a + ib$, $a - ib$, or $-a + ib$, $-a - ib$. However, the wave patterns associated with the first pair are drastically different from those corresponding to the second pair. Indeed, multiple roots of dispersion relations of RW systems, e.g., long-wave–short-wave interaction model [30], are known to produce different wave profiles.

### III. CONDITIONS FOR THE EXISTENCE OF ROGUE WAVES AND THE CORRELATION WITH MODULATION INSTABILITY

#### A. The existence condition for RWs

The solution given by Eqs. (2) and (5) is a RW only for $b \equiv \text{Im}\{\Omega_0\} \neq 0$, otherwise the result is just a plane wave. It will be instructive to review the situation for the single-component DNLSE (with real $\sigma_0$ and $\gamma_0$):

$$iA_t + A_{xx} - \sigma_0|A|^2 A + i\gamma_0|A|^2 A_x = 0. \tag{8a}$$

Our works earlier had established the existence criterion as [25]

$$\sigma_0 < \frac{\gamma_0 \rho_0^2}{4}, \tag{8b}$$

where $\rho_0$ is the amplitude of the plane wave. Remarkably, RWs can exist even if the cubic nonlinearity is self-defocusing ($\sigma_0 > 0$ in Eq. (8a)), provided that a sufficiently strong self-steepening term (as measured by $\gamma_0$) is present. In the absence of self-steepening ($\gamma_0 = 0$), one recovers the known condition for NLS ($\sigma_0 < 0$ for the existence of rogue waves).

We can now turn to the dispersion relation of the fully coupled system (Eq. (7)):

*The self-focusing case* ($\sigma < 0$)

In this case, Eq. (7) gives rise to one pair of complex conjugate roots if $\delta = 0$, and to two pairs of such roots for $\delta \neq 0$. Consequently, RWs always exist.

*The self-defocusing case* ($\sigma > 0$)

Naturally the criterion for the existence of RWs is more restrictive in this case. By evaluating the discriminant of Eq. (7) and assuming $\delta \neq 0$, we conclude that RWs can occur if either

$$S < \frac{\Gamma^2}{2}, \quad \text{or} \tag{9a}$$

$$S > R(\delta, \Gamma) = \frac{6\Gamma^2 - 5\delta^2}{4} + \xi + \frac{9\delta^2(\delta^2 - 8\Gamma^2)}{16\xi}, \tag{9b}$$

where

$$\xi^3 \equiv \frac{27}{64}\delta^2\left(\delta^4 + 20\Gamma^2\delta^2 - 8\Gamma^4\right) + \frac{27}{8}\delta^2\sqrt{\Gamma^2\left(\Gamma^2 + \delta^2\right)^3}.$$

For the special case of $A = B$, $\delta = 0$, Eq. (1) degenerates into a single component DNLSE with $\sigma_0 = 2\sigma$, $\gamma_0 = 2\gamma$, and Eq. (8b) is then equivalent to Eq. (9a).

For the case defined as per Eq. (9a) and $\delta \neq 0$, there are two pairs of complex conjugate roots, with the real part of one pair being of opposite sign to that of the other pair. However, the two respective RWs exhibit different configurations, as discussed below in Section IV. In the case defined as per Eq. (9b), there exists strictly one pair of complex conjugate roots. In other words, the number of RWs depends on the magnitude of the cubic nonlinearity.

The effect of coupling on the existence condition for RWs of Eq. (1) can now be highlighted:

● Complex roots of the dispersion relation (Eq. (7)) are necessary for RWs to exist. For the case of Eq. (9b), equal real roots become complex as $\delta$ starts to increase from zero to a positive value (Table 1). Physically, the mismatch in group velocities of the two components generates these RWs. Consequently, coupling extends the existence range of the RWs tremendously by incorporating the region of $S > R$ (see Eq. (9b)).

- In the case of $0 \leq S < \Gamma^2/2$, which also holds for the uncoupled equations, an extra pair of complex roots produces a different RW for the same set of input parameters. This coexistence of distinct RWs for the same input parameters of the system in the presence of the coupling is also observed in other coupled evolution equations [30].

We now aim to identify conditions imposed on the cubic nonlinearity which are necessary to sustain the existence of RWs, if one varies either the group-velocity mismatch ($\delta$) or self-steepening ($\gamma$, or $\Gamma$ in Eq. (4)), and keep the other parameters constant.

*Varying $\delta$ at constant $\Gamma$*

With constant $\Gamma$, the constraint $0 \leq S < \Gamma^2/2$ becomes a horizontal straight line, with two distinct RWs existing below this line (Fig. 1). The upper curve in Fig. 1 represents the condition $S = R(\delta, \Gamma)$, see Eq. (9b). The dependence on $\delta$ is not monotonic. For $2\delta < \Gamma$, $R$ decreases as $\delta$ increases. The turning point is attained at $\delta = \Gamma/2$ with $R = \Gamma^2$. Further increase of $\delta$ will require a larger cubic nonlinearity to start a RW. Physically this is meaningful too, as an indefinitely large walkoff between the two components suppresses the interaction between them.

*Varying Γ at constant δ*

For a fixed group-velocity mismatch (δ), the increase of the self-steepening parameter |Γ| allows a larger range of the strength of the cubic nonlinearity to sustain a RW by means of the single-mode mechanism (the region below the solid curve in Fig. 2). However, the increase in absolute value of the self-steepening parameter implies that the coupling becomes less effective in producing RWs, as the region above the dotted curve shrinks in Fig. 2.

For Γ = 0, Eq. (1) reduces to the self-defocusing Manakov system, and Eq. (9b) simplifies to

$$4S > \delta^2 \qquad (10)$$

as the only regime admitting the existence of one rogue wave. This is consistent with the results known for the defocusing Manakov system, which admits MI and RWs [13, 31], under constraints similar to that given by Eq. (10). The system of coupled DNLSEs combines the existence regimes of the single component DNLSE and the Manakov system, allowing RWs under both small and large regimes of cubic nonlinearity.

### B. Modulation instability

To analyze MI, we consider small wavy perturbations $\exp[i(rx - st)]$ imposed onto the plane wave solution, $A = \rho \exp(i\beta x - i\omega_1 t)$, $B = \rho \exp(i\beta x - i\omega_2 t)$, with β,

$\omega_1$ and $\omega_2$ given by Eq. (2). The dispersion relation for the perturbations is derived as

$$s^4 - 2(2\sigma\rho^2 + \delta^2 + r^2)r^2 s^2 + 4\gamma\rho^2(\gamma^2\rho^4 - 2\sigma\rho^2)r^3 s$$
$$+ (\gamma^2\rho^4 + \delta^2 - r^2)(3\gamma^2\rho^4 - 4\sigma\rho^2 + \delta^2 - r^2)r^4 = 0.$$

With $c \equiv s/r = O(1)$ and taking the long wavelength limit of $r \to 0$, MI occurs for low frequencies and wave numbers (small $s$ and $r$, which is the 'baseband' MI [13, 14]) if there are complex roots for the equation

$$c^4 - 2(2\sigma\rho^2 + \delta^2)c^2 + 4\gamma\rho^2(\gamma^2\rho^4 - 2\sigma\rho^2)c + (\gamma^2\rho^4 + \delta^2)(3\gamma^2\rho^4 - 4\sigma\rho^2 + \delta^2) = 0. \quad (11)$$

Remarkably Eq. (11) is identical to the dispersion relation of the breather modes in the long wavelength limit, as given by Eqs. (4) and (7). This intimate connection between the RWs and zero frequency MI has been recognized for many evolution equations, e.g., NLSE [32], coupled NLSEs [13], and DNLSE [25]. This trend is thus confirmed again here for the present system of coupled DNLSEs.

A typical MI gain spectrum versus the perturbation wavenumber $r$ is displayed in Fig. 3. Although additional bands may be generated, in general only the long-wave / low-frequency portion is critical for the consideration of the RWs. A more striking example is provided in Fig. 4, where the gain band for a sufficiently large group velocity mismatch does not possess a small-$r$ limit, and hence no RW exists, in agreement with the analysis presented in the previous section.

## IV. WAVE PROFILES AND DYNAMICS

An obviously important characteristic of a RW is its amplitude. For the NLSE, the *amplification ratio* (the largest displacement featured by a RW solution divided by that of the background) for the Peregrine breather is 3 [5, 6], which also holds for the first-order RW solutions of the DNLSE [25]. For the integrable Manakov system, the amplification ratio of first order RWs without group velocity mismatch *cannot* exceed three [10, 11, 13]. Remarkably, first order RWs of the present system may have an amplification ratio *greater than three*.

*Conservation of the total norm*

We begin the consideration of the dynamics by noting that Eqs. (1) exhibit the conservation of the total norm, $\frac{d}{dt}\left[\int_{-\infty}^{\infty}(|A|^2+|B|^2)dx\right]=0$. To comply with this conservation law, elevations in the wave profile must be accompanied by depressions.

*The amplification ratio exceeding three*

First we address the case of zero group velocity mismatch ($\delta = 0$). Wave patterns similar to those produced by the Manakov system and single component

DNLSE are observed in this case, namely elevation RWs (Fig. 5). Their amplification ratio is three. However, in the presence of the group velocity mismatch, e.g. $\delta = 0.5$, the largest amplitude may exceed the background level by a factor of $\approx 3.9$ (Fig. 6a).

*Multiple roots of the dispersion relation*

For the regime where $S < \Gamma^2/2$ and $\delta \neq 0$, there are two possible RW modes corresponding to the two pairs of complex roots of the dispersion relation. Figure 7 displays a wave profile for input parameters identical to those in Fig. 6, except for a different $\Omega_0$. For this configuration, both components of Eq. (1) feature four-petal configurations. The RW pattern displayed in Fig. 7 has an essentially larger spatial extent than its counterpart in Fig. 6.

*Variation in wave profiles*

For the classical NLSE, the only possible shape for the first order RW is an elevation mode, but RWs for the Manakov system can also exhibit depression and four-petal configurations. A similar scenario prevails in the present DNLSE system. For the single component DNLSE, leading order RWs in the form of elevation patterns were documented earlier [25]. For coupled DNLSEs, depression and four-petal patterns appear, see Figs. 6 and 7. Profiles of the RWs in this regime follow these familiar patterns, see Fig. 8.

## V. GAUGE TRANSFORMATIONS AND APPLICATIONS

Gauge transformations will be utilized to illustrate additional realizations of the system of coupled DNLSEs (Eq. (1)). First we consider the special case of $\delta = 0$,

$$iA_t + A_{xx} - \sigma(|A|^2 + |B|^2)A + i\gamma(A^*A_x + B^*B_x)A = 0,$$

$$iB_t + B_{xx} - \sigma(|A|^2 + |B|^2)B + i\gamma(A^*A_x + B^*B_x)B = 0, \quad (12)$$

which was studied earlier [26, 33]. The gauge transformation [26, 33],

$$P = A\exp\left[-\frac{i\gamma}{2}\int(|A|^2 + |B|^2)\,dx\right], \quad Q = B\exp\left[-\frac{i\gamma}{2}\int(|A|^2 + |B|^2)\,dx\right],$$

will transform Eq. (12) to the evolution system

$$iP_t + P_{xx} - \sigma(|P|^2 + |Q|^2)P + i\gamma\left[(|P|^2 + |Q|^2)P\right]_x = 0,$$

$$iQ_t + Q_{xx} - \sigma(|P|^2 + |Q|^2)Q + i\gamma\left[(|P|^2 + |Q|^2)Q\right]_x = 0. \quad (13)$$

The gauge transformation for $\delta \neq 0$ will be considered elsewhere. Eq. (13) is then recognized as the governing system for the evolution of slowly varying wave amplitudes investigated earlier in optics and plasmas:

● In optics, *t* and *x* are the propagation distance and retarded time [12] in terms of fiber waveguides. For ultrashort pulses, self-steepening effects (alias *shocking*) should be taken into account. These effects are incorporated in Eq. (13) through terms involving the parameter γ [19, 34, 35]. A consistent derivation of models of this type for the propagation of bimodal (two-polarization) narrow optical pulses in fibers from the Maxwell's equations for a medium with the Kerr nonlinearity is available [34, 35]. In particular, the equality of the SPM and XPM coefficients, as adopted in Eq. (13), is possible for a special elliptic birefringence in the fiber [36]. In this case, the cubic nonlinearity is self-focusing, i.e., σ is negative in Eq. (13). The relative size of the shock term, in comparison with the usual Kerr terms, is estimated as $1/(T_0\omega_0)$, where $T_0$ is the temporal duration of the pulse, and $\omega_0$ is the carrier frequency (typically about $2\pi \times 100$ THz). For the usual picosecond pulses, the shock terms are completely negligible. However, for few-cycle or sufficiently short pulses, they play an important role, attaining a relative magnitude of about 0.1 [12]. In particular, conspicuous self-steepening was observed in the course of compressing a pulse from 40 down to 8 fs [37]. However, a consistent model for the transmission of such short pulses must also include additional higher-order terms, such as the third-order group-velocity dispersion and the Raman term [1]. Hence the present model may not be of adequate accuracy in terms of fiber optics.

● *On the contrary, the same model plays a fundamental role in plasma physics*. In warm multi-species plasmas with anisotropic pressures and different equilibrium drifts, coupled DNLSEs govern the oblique propagation of nonlinear magnetohydrodynamic waves relative to an external magnetic field under certain conditions, dropping the quasi-neutrality assumption [20]. If Alfvén waves in a magnetized plasma bear both right and left circular polarizations, the equations for the respective amplitudes lead to the system of DNLSEs in the form of Eq. (1). In this context, the model for the dynamics of Alfvén waves, based on Eqs. (12) and (13), is a fundamental one, as it does not require the addition of any higher-order terms in the physically relevant situation.

## VI. NUMERICAL STUDIES ON THE EFFECTS OF PERTURBATIONS

In physical and engineering realizations of RWs, e.g., oceanic surface waves, there is usually a weakly chaotic background, the robustness of the RWs in such an environment will be an important issue. It is thus necessary to test this aspect of the system through numerical simulations. For this purpose, a split-step Fourier algorithm was applied for the simulations of Eq. (1). For a weak noise level of 1%, the configurations of both the 'elevation-elevation' and 'elevation - four-petal' types exhibit reasonable robustness, in the sense that the growth phase of the RWs persists in an essentially unaffected form on top of only a mildly disturbed background (Figs. 9 and 10).

However, if a relatively strong noise level of 5% is imposed, the growth phase of the RW is completely 'masked' by the instability of the background (Fig. 11).

## VII. CONCLUSION

Coupled multi-wave systems may produce new MI (modulation instability) regimes and RW (rogue wave) patterns. For coupled NLSEs, this scenario arises even if nonlinearity and dispersion are of opposite signs in each component [12]. However, only MI bands which include the zero frequency limit will generate RWs [13]. The main goal of the present work is to extend the study to coupled DNLSEs. In addition to the constraints for the existence of RWs supported by individual components, additional MI regimes and RW modes are identified for such coupled DNLSEs. The role of the physical parameters, namely those controlling the group velocity mismatch, cubic nonlinearity and self-steepening, was investigated. The effect of the group velocity mismatch parameter is thus considerably more complicated than its corresponding role in the coupled NLS case [31]. In the latter case, the mismatch is proportional to the background amplitude. In the present DNLSE system, not only is the dependence on the background amplitude more involved, but a difference in the group velocity actually introduces new RW regimes.

Remarkable features of the newly found RW modes include a higher than normal amplification ratio and transformations of wave profiles. More precisely, the amplification ratio of the first order RWs for the single component NLSE and DNLSE, as well as the coupled NLSEs without group velocity mismatch, does not exceed three. On the other hand, for the present system of coupled DNLSEs, configurations with amplification ratio as high as 3.9 can be found in a suitable range of the group velocity mismatch. Although it is known that the amplification ratio can exceed three for the single component NLSE with a 'fluctuating' background [38], the present result is still a very remarkable one for first order RWs built on top of a constant background. Regarding transformations of wave profiles, 'four-petal' configurations are known to result from the splitting of peaks or valleys of elevation type RWs by varying the relative frequency in systems of coupled NLSEs [39]. We have demonstrated here that a change in the group velocity mismatch can induce similar profile transformation for RWs in coupled DNLSE systems. In this perspective, it is relevant to note that the change of the relative frequency is linked to a mismatch in the group velocities. The connection with long-wave models [40] has been examined too.

By using gauge transformations, the present system of DNLSEs can be mapped into evolution models which have direct applications to optics and plasmas. For plasma physics, the Kaup-Newell form of the coupled DNLSEs, represented by Eq. (13), provides a fundamental model for the evolution of Alfvén wave packets [20, 21]. On the contrary, in terms of fiber optics, the self-steepening nonlinear term in the DNLSEs is only one correction of a higher order NLSE, which should be introduced for modeling the transmission of short pulses. The parameter range for optical fibers where such self-steepening RW patterns may be expected was estimated, but, unlike the dynamical model for the Alfvén waves in plasmas, the DNLSE system in the present form (without additional terms) is not expected to be an accurate model in fiber optics. To analyze this issue further, we have performed additional simulations of the DNLSE system, with each equation including the term accounting for the intra-pulse Raman effect. The result (not shown here in detail) is that, although such an extended model does not admit analytical solutions, the numerically generated RWs are quite similar to those reported in this paper, for realistic values of the strength of the Raman terms.

Many additional questions remain to be addressed. An obvious one is the search for higher order rogue waves, with a particular aim to find still higher values of the amplification ratio and new wave profiles. This aim is suggested by the known fact that the second order RWs for the NLSEs will have a higher amplitude than the first order ones. A very relevant issue, but seldom addressed in the literature, is a numerical test for stability for such higher order rogue waves (which should be decoupled from the inevitable MI of the flat background). Furthermore, scenarios with distinct background amplitudes for each of the different components remain to be investigated. Finally, extension to the case of variable coefficient DNLSEs, with applications to inhomogeneous media, can be pursued too.

## ACKNOWLEDGEMENTS

Partial financial support for this project has been provided by the Research Grants Council contracts HKU711713E and HKU17200815. B.A.M. appreciates hospitality of the Department of Mechanical Engineering at the University of Hong Kong.

**Table Caption**

Table 1: The classification scheme of roots of Eq. (7) for different values of the group velocity mismatch parameter $\delta$

**Figures Captions**

(1) FIG. 1. The existence region of RW (rogue wave) in the plane of parameters $S$ (the strength of the defocusing cubic nonlinearity, see Eq. (4)) and $\delta$ (group-velocity mismatch) for $\gamma = 2$ and $\rho = 1$. Above the dashed curve ($S = R$) and below the solid line ($S = \Gamma^2/2$), RWs exist. Between the dashed and solid boundaries, RWs do not exist.

(2) FIG. 2. The existence region of RW (rogue wave) in the plane of $S$ (the strength of the defocusing cubic nonlinearity, see Eq. (4)) and $\Gamma$ (the strength of the self-steepening nonlinearity) for $\delta = 1$ and $\rho = 1$. Above the dashed curve ($S = R$) and below the solid line ($S = \Gamma^2/2$), RWs exist. Between the dashed and solid curves, RWs do not exist.

(3) FIG. 3. The spectrum of modulation instability gain for $S = 0.1$, $\Gamma = 1$, and $\delta = 0, 0.5, 1$.

(4) FIG. 4. The spectrum of modulation instability gain for $S = 1.1$, $\Gamma = 1$, $\delta = 0.5$ and $1$.

(5) FIG. 5. An example of rogue waves of the 'elevation-elevation' type for $\rho = 1$, $\sigma = 0.1$, $\gamma = 1$, $\delta = 0$, $\Omega_0 = 1 + 1.26i$, (a) $|A|$ and (b) $|B|$ versus $x$ and $t$.

(6) FIG. 6. An example of rogue waves of the 'elevation - four-petal' type for $\rho = 1$, $\sigma = 0.1$, $\gamma = 1$, $\delta = 0.5$, $\Omega_0 = 1.11 + 1.23i$, (a) $|A|$ and (b) $|B|$ versus $x$ and $t$.

(7) FIG. 7. A rogue wave with the same input parameters as those of Fig. 6 except for a different complex frequency $\Omega_0 = -1.11 + 0.255i$, (a) $|A|$ and (b) $|B|$ versus $x$ and $t$.

(8) FIG. 8. An example of a rogue wave in the additional regime of existence due to coupling, $\rho = 1$, $\sigma = 1.1$, $\gamma = 1$, $\delta = 0.5$, $\Omega_0 = -0.542 + 0.243i$, (a) $|A|$ and (b) $|B|$ versus $x$ and $t$.

(9) FIG. 9. Growth phase of a rogue wave versus $x$ and $t$ ($t$ from -4 to 0) is robust against a sufficiently small perturbation of 1%, with parameters the same as those of Figure 5. Top: Exact solution; Bottom: Evolution in the presence of a 1% noise.

(10) FIG. 10. Growth phase of another rogue wave versus $x$ and $t$ ($t$ from -4 to 0) is also robust against a sufficiently small perturbation of 1%, with parameters the same as those of Figure 6. Top: Exact solution; Bottom: Evolution in the presence of a 1% noise.

(11) FIG. 11. The growth phase of the rogue wave versus $x$ and $t$ ($t$ from -3 to 0) is completely masked by the modulation instability of the background if the noise is sufficiently strong (5% here), with parameters the same as those of Figure 6. Top: Exact solution; Bottom: Evolution in the presence of a 5% noise.

|  | $\delta = 0$ | $\delta > 0$ |
| --- | --- | --- |
| $0 \leq S < \dfrac{\Gamma^2}{2}$ | 1 pair of complex-conjugate roots and a double real root | 2 pairs of complex-conjugate roots |
| $\dfrac{\Gamma^2}{2} \leq S \leq R$ | 4 real roots | 4 real roots |
| $S > R$ | 2 different real roots and 2 equal real roots | 1 pair of complex- conjugate roots and 2 real roots |

Table 1: The classification scheme of roots of Eq. (7) for different values of the group velocity mismatch parameter $\delta$

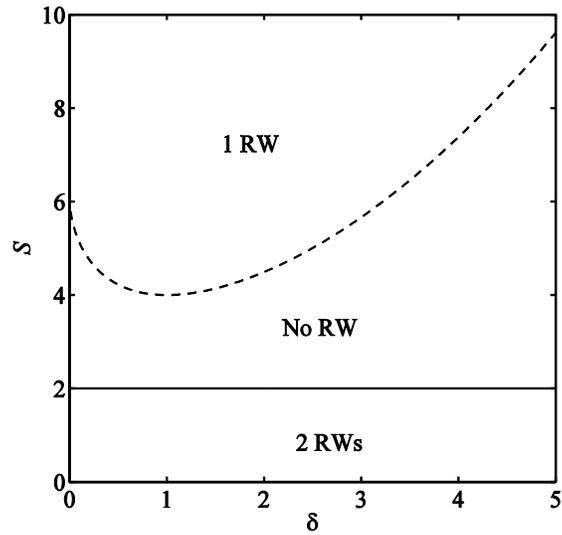

**Figure 1: The existence region of RW (rogue wave) in the plane of parameters $S$ (the strength of the defocusing cubic nonlinearity, see Eq. (4)) and $\delta$ (group-velocity mismatch) for $\gamma = 2$ and $\rho = 1$. Above the dashed curve ($S = R$) and below the solid line ($S = \Gamma^2/2$), RWs exist. Between the dashed and solid boundaries, RWs do not exist.**

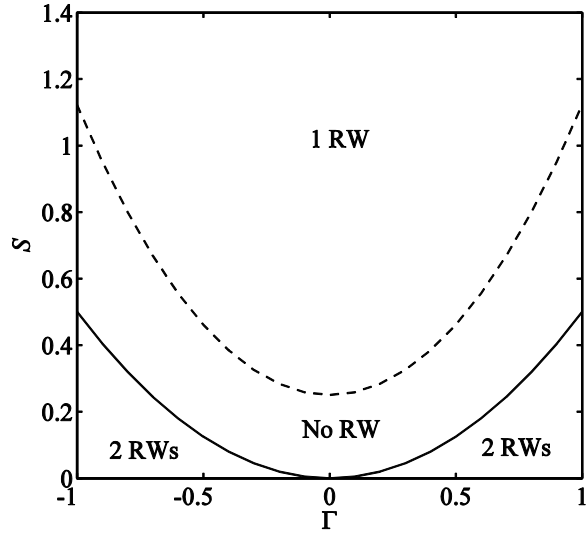

**Figure 2: The existence region of RW (rogue wave) in the plane of $S$ (the strength of the defocusing cubic nonlinearity, see Eq. (4)) and $\Gamma$ (the strength of the self-steepening nonlinearity) for $\delta = 1$ and $\rho = 1$. Above the dashed curve ($S = R$) and below the solid line ($S = \Gamma^2/2$), RWs exist. Between the dashed and solid curves, RWs do not exist.**

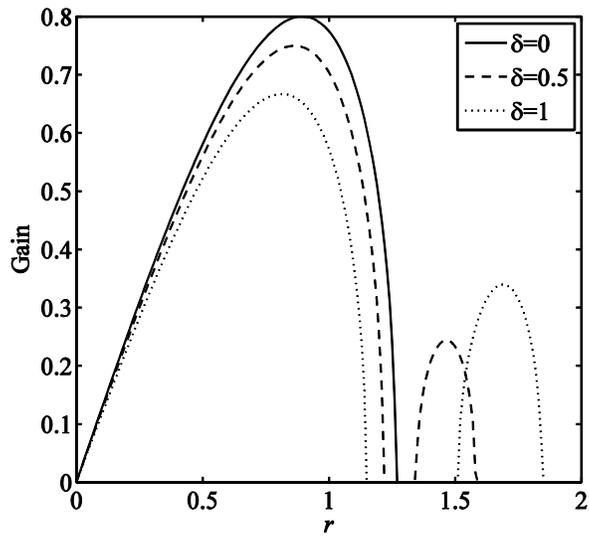

**Figure 3: The spectrum of modulation instability gain for *S* = 0.1, Γ = 1, and δ = 0, 0.5, 1.**

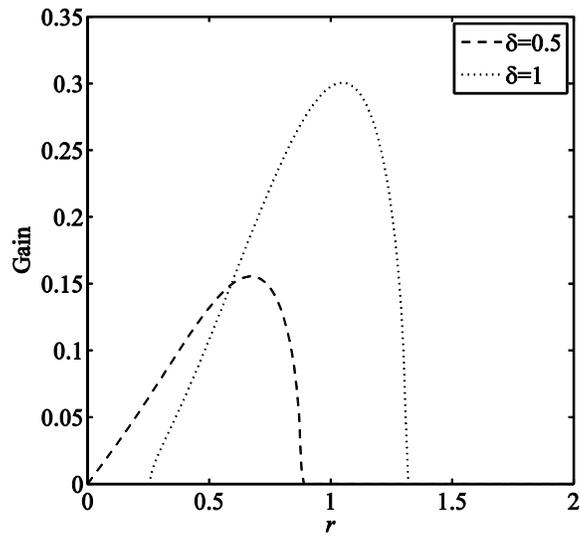

**Figure 4: The spectrum of modulation instability gain for *S* = 1.1, Γ = 1, δ = 0.5 and 1**.

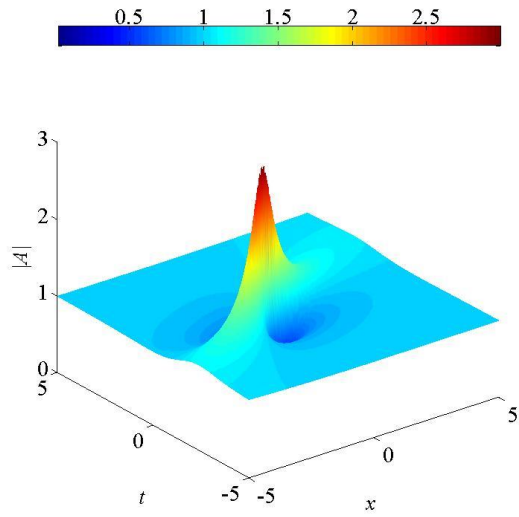

(a)

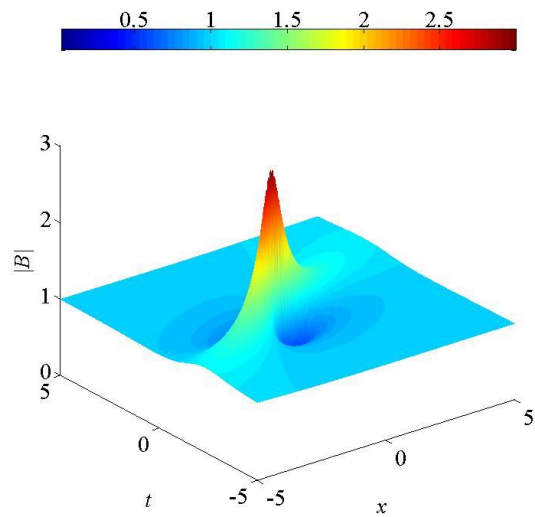

(b)

**Figure 5: An example of rogue waves of the 'elevation-elevation' type for $\rho = 1$, $\sigma = 0.1$, $\gamma = 1$, $\delta = 0$, $\Omega_0 = 1 + 1.26i$, (a) $|A|$ and (b) $|B|$ versus $x$ and $t$.**

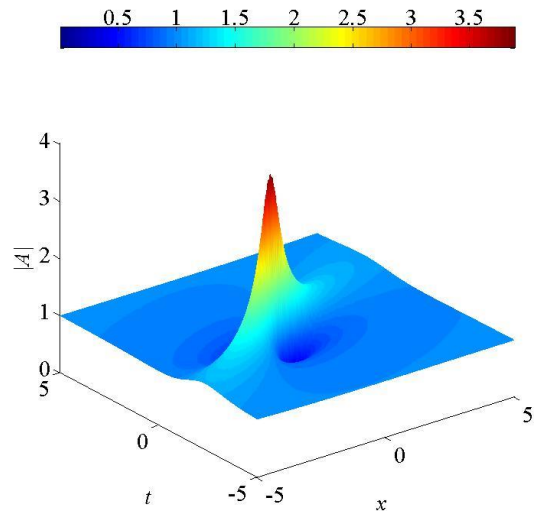

(a)

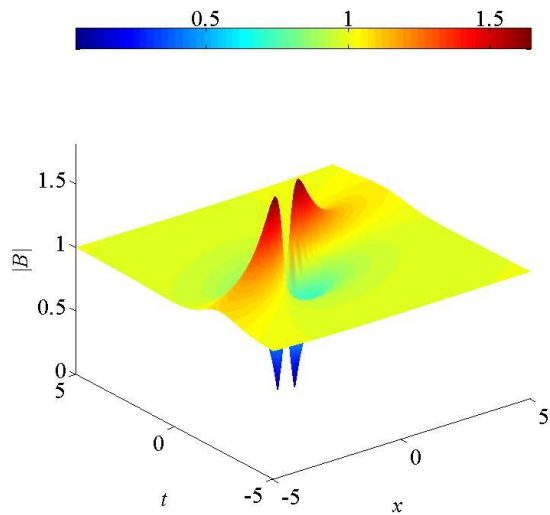

(b)

**Figure 6: An example of rogue waves of the 'elevation - four-petal' type for ρ = 1, σ = 0.1, γ = 1, δ = 0.5, $\Omega_0$ = 1.11 + 1.23*i*, (a) |*A*| and (b) |*B*| versus *x* and *t*.**

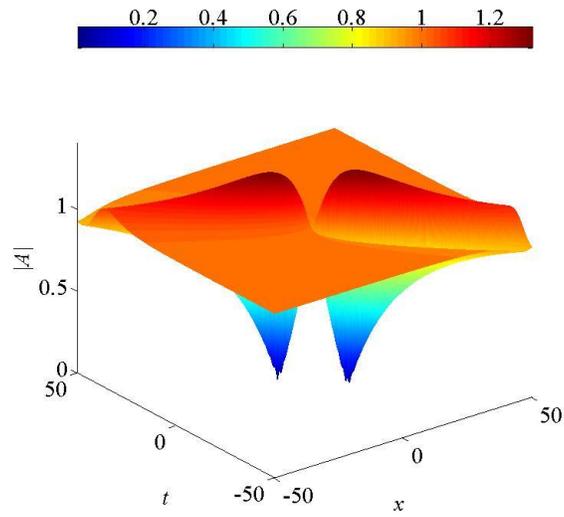

(a)

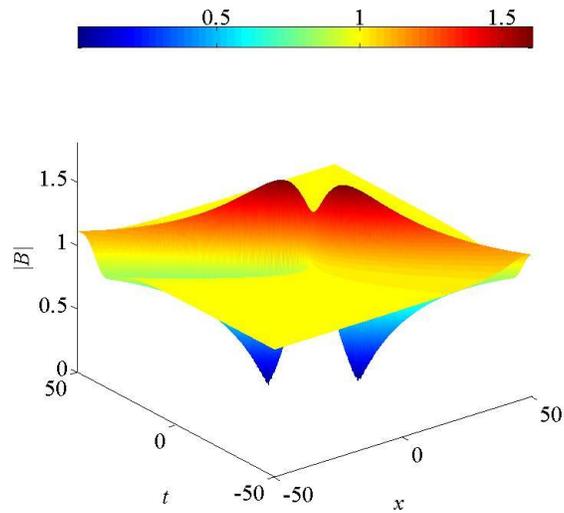

(b)

**Figure 7: A rogue wave with the same input parameters as those of Fig. 6 except for a different complex frequency $\Omega_0 = -1.11 + 0.255i$, (a) |A| and (b) |B| versus *x* and *t*.**

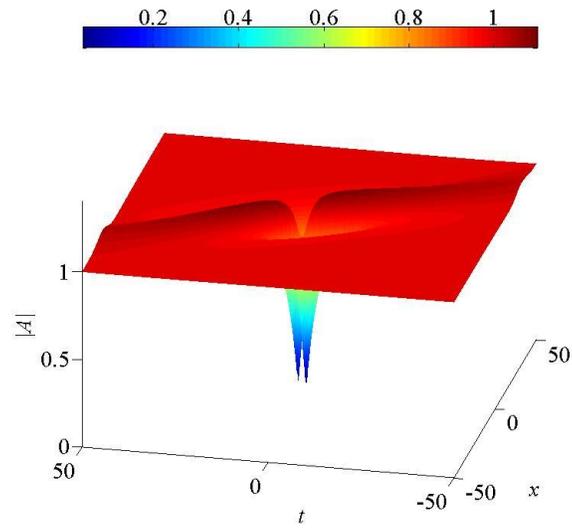

(a)

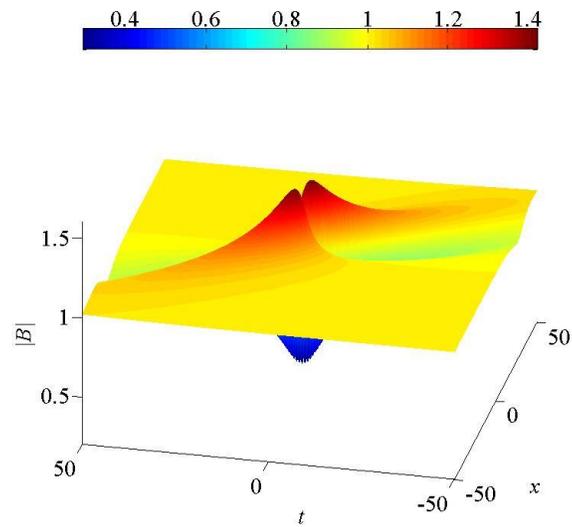

(b)

**Figure 8: An example of a rogue wave in the additional regime of existence due to coupling, ρ = 1, σ = 1.1, γ = 1, δ = 0.5, $\Omega_0$ = –0.542 + 0.243*i*, (a) |*A*| and (b) |*B*| versus *x* and *t*.**

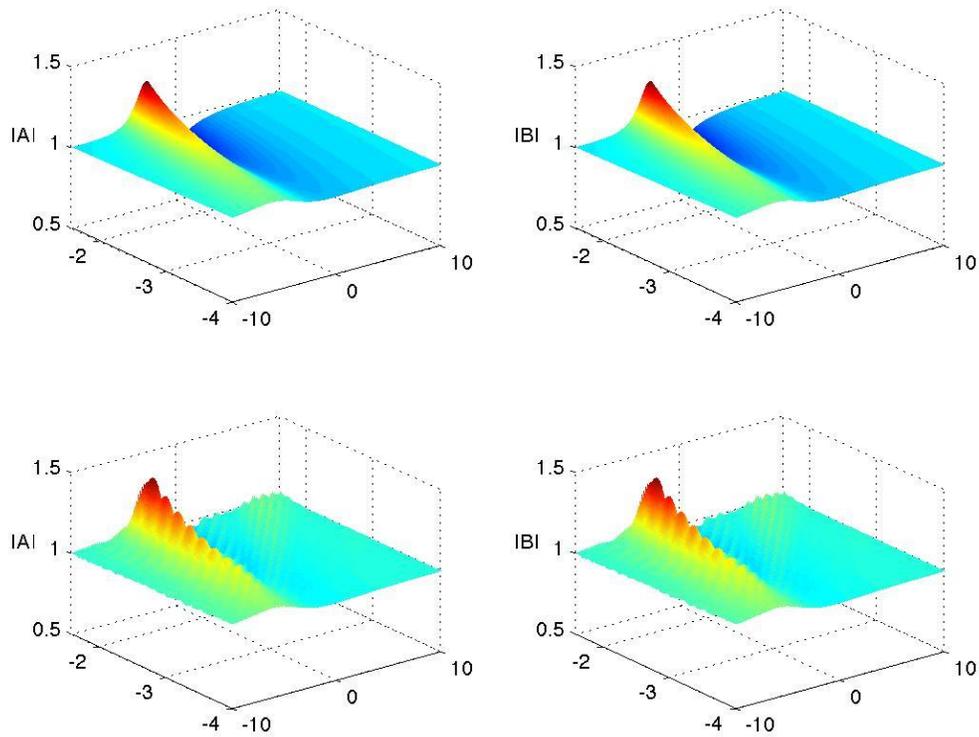

**Figure 9: Growth phase of a rogue wave (versus *x* and *t*, *t* from -4 to 0) is robust against a sufficiently small perturbation of 1%, with parameters the same as those of Figure 5. Top: Exact solution; Bottom: Evolution in the presence of a 1% noise.**

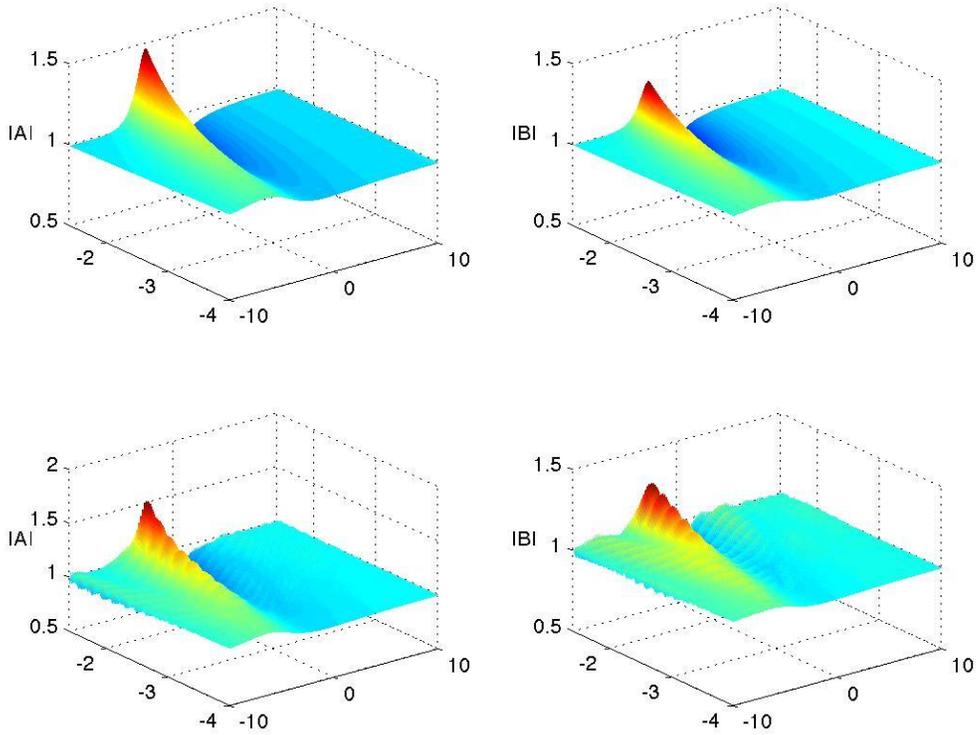

**Figure 10: Growth phase of another rogue wave (versus *x* and *t*, *t* from -4 to 0) is also robust against a sufficiently small perturbation of 1%, with parameters the same as those of Figure 6. Top: Exact solution; Bottom: Evolution in the presence of a 1% noise.**

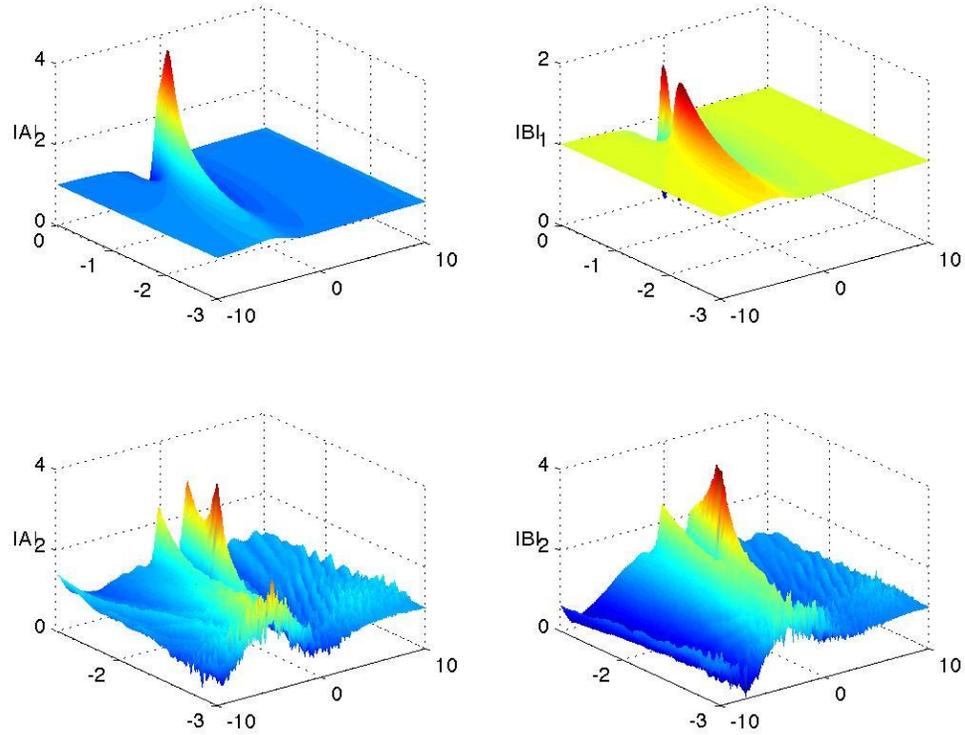

**Figure 11: The growth phase of the rogue wave versus ($x$ and $t$, $t$ from -3 to 0) is completely masked by the modulation instability of the background if the noise is sufficiently strong (5% here), with parameters the same as those of Figure 6. Top: Exact solution; Bottom: Evolution in the presence of a 5% noise.**